\documentclass[twocolumn]{aastex631}

\usepackage{color, booktabs, subfigure, multirow}
\usepackage{url, hyperref}
\usepackage[capitalise]{cleveref}

\shorttitle{X-ray polarization from BL Lacertae}
\shortauthors{Peirson et al.}

\graphicspath{{./}{figures/}}

\begin{document}

\title{X-ray Polarization of BL Lacertae in Outburst}

\correspondingauthor{Abel L. Peirson}
\email{alpv95@stanford.edu}
\author[0000-0001-6292-1911]{Abel L. Peirson}
\affiliation{Department of Physics and Kavli Institute for Particle Astrophysics and Cosmology, Stanford University, Stanford, California 94305, USA}

\author[0000-0002-6548-5622]{Michela Negro}
\affiliation{University of Maryland, Baltimore County, Baltimore, MD 21250, USA}
\affiliation{NASA Goddard Space Flight Center, Greenbelt, MD 20771, USA}
\affiliation{Center for Research and Exploration in Space Science and Technology, NASA/GSFC, Greenbelt, MD 20771, USA}

\author[0000-0001-9200-4006]{Ioannis Liodakis}
\affiliation{Finnish Centre for Astronomy with ESO, 20014 University of Turku, Finland}

\author[0000-0001-9815-9092]{Riccardo Middei}
\affiliation{Space Science Data Center, Agenzia Spaziale Italiana, Via del Politecnico snc, 00133 Roma, Italy}
\affiliation{INAF Osservatorio Astronomico di Roma, Via Frascati 33, 00078 Monte Porzio Catone (RM), Italy}

\author[0000-0001-5717-3736]{Dawoon E. Kim}
\affiliation{INAF Istituto di Astrofisica e Planetologia Spaziali, Via del Fosso del Cavaliere 100, I-00133 Roma, Italy}
\affiliation{Dipartimento di Fisica, Università degli Studi di Roma “La Sapienza”, Piazzale Aldo Moro 5, I-00185 Roma, Italy}
\affiliation{Dipartimento di Fisica, Università degli Studi di Roma “Tor Vergata”, Via della Ricerca Scientifica 1, I-00133 Roma, Italy}

\author[0000-0001-7396-3332]{Alan P. Marscher}
\affiliation{Institute for Astrophysical Research, Boston University, 725 Commonwealth Avenue, Boston, MA 02215, USA}

\author[0000-0002-6492-1293]{Herman L. Marshall}
\affiliation{MIT Kavli Institute for Astrophysics and Space Research, Massachusetts Institute of Technology, 77 Massachusetts Avenue, Cambridge, MA 02139, USA}

\author[0000-0001-6897-5996]{Luigi Pacciani}
\affiliation{INAF Istituto di Astrofisica e Planetologia Spaziali, Via del Fosso del Cavaliere 100, 00133 Roma, Italy}

\author[0000-0001-6711-3286]{Roger W. Romani}
\affiliation{Department of Physics and Kavli Institute for Particle Astrophysics and Cosmology, Stanford University, Stanford, California 94305, USA}

\author[0000-0002-7568-8765]{Kinwah Wu}
\affiliation{Mullard Space Science Laboratory, University College London, Holmbury St Mary, Dorking, Surrey RH5 6NT, UK}

% ...

\begin{abstract}
We report the first $> 99\%$ confidence detection of X-ray polarization in BL Lacertae. During a recent  X-ray/$\gamma$-ray outburst, a 287 ksec observation (2022 November 27-30) was taken using the Imaging X-ray Polarimetry Explorer ({\it IXPE}), together with contemporaneous multiwavelength observations from the Neil Gehrels {\it Swift} observatory and {\it XMM-Newton} in soft X-rays (0.3--10~keV), {\it NuSTAR} in hard X-rays (3--70~keV), and optical polarization from the Calar Alto, 
and Perkins Telescope observatories. Our contemporaneous X-ray data suggest that the {\it IXPE} energy band is at the crossover between the low- and high-frequency blazar emission humps. The source displays significant variability during the observation, and we measure polarization in three separate time bins. Contemporaneous X-ray spectra allow us to determine the relative contribution from each emission hump. We find $>99\%$ confidence X-ray polarization $\Pi_{2-4{\rm keV}} = 21.7^{+5.6}_{-7.9}\%$ and electric vector polarization angle $\psi_{2-4{\rm keV}} = -28.7 \pm 8.7^{\circ}$ in the time bin with highest estimated synchrotron flux contribution. We discuss possible implications of our observations, including previous {\it IXPE} BL Lacertae pointings, tentatively concluding that synchrotron self-Compton emission dominates over hadronic emission processes during the observed epochs.

\end{abstract}


\section{Introduction} \label{sec:intro}

Blazars are active galactic nuclei (AGN) that launch collimated relativistic jets of plasma oriented within a few degrees from the observer's line of sight \citep[and references therein]{blandford_relativistic_2019}. The Doppler-boosted jet emission dominates the observed spectral energy distribution (SED) that extends from radio to $\gamma$-rays and is characterized by two emission components. Blazars are often classified by the peak frequency of the low-energy (electron synchrotron radiation) component  as low-synchrotron peaked (LSP, $\nu_{\rm syn}<10^{14}~{\rm Hz}$), intermediate-synchrotron peaked (ISP $10^{14}<\nu_{\rm syn}<10^{15}~{\rm Hz}$), and high-synchrotron peaked (HSP $\nu_{\rm syn}>10^{15}~{\rm Hz}$, \citealp{abdo_spectral_2010}). Here we focus on ISPs and in particular BL Lacertae (BL Lac), whose recent $\gamma$-ray outburst (see Appendix \ref{sec:app_multi}) briefly boosted its characteristic peak energies, moving it from the LSP into the ISP class.

\begin{figure}[t]
  \centering
  \includegraphics[width=1\linewidth]{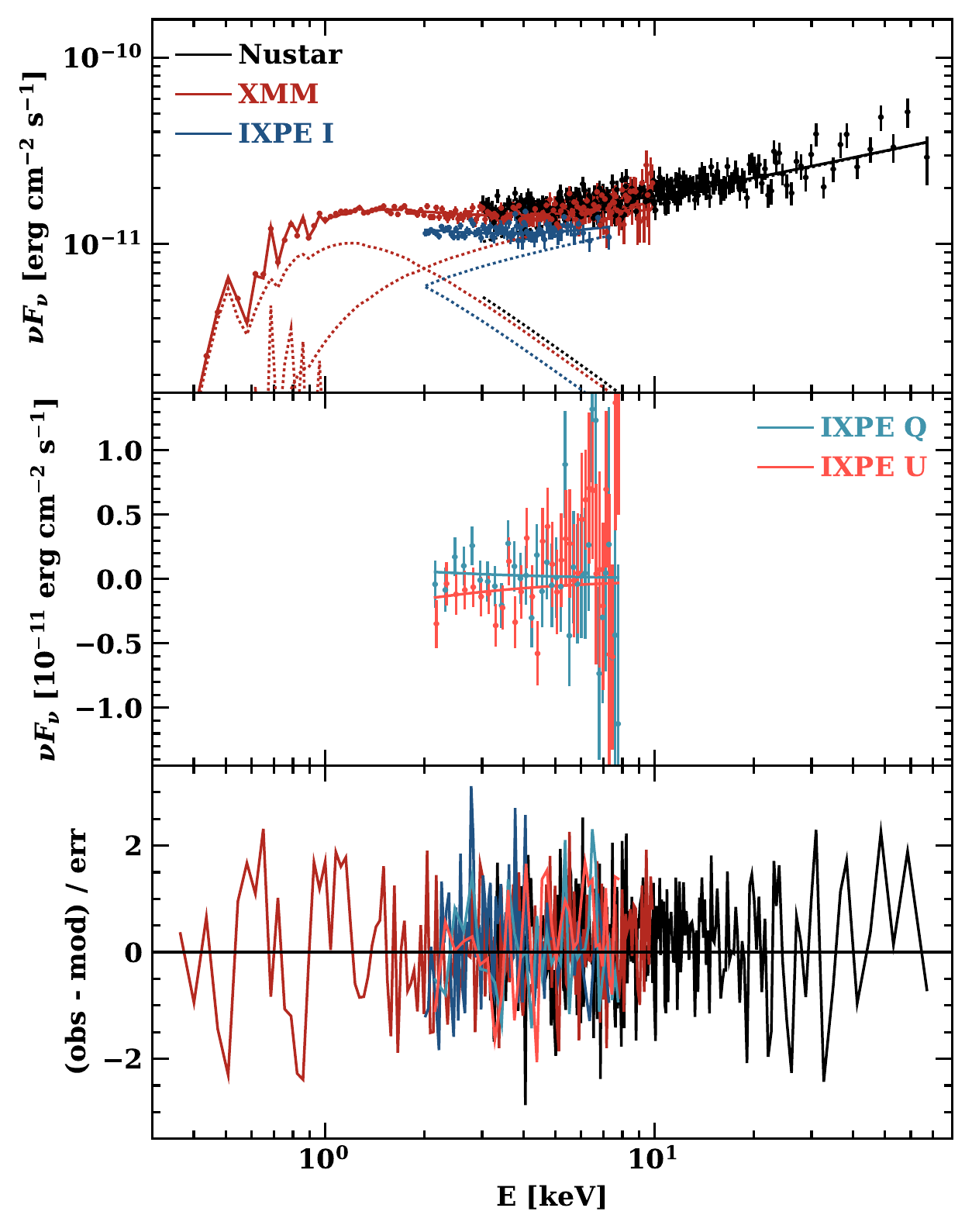}
  \caption{Quasi-simultaneous X-ray spectra during the {\it IXPE} observation. The fitted model is the sum of two absorbed power-laws and an apec component. Dotted lines in the top panel show the individual power-law and apec components. The middle panel shows the polarized spectra with constant polarization for the low-energy component; the high energy component has fixed polarization degree PD=0.  The significance of the spectral fit residuals is shown in the bottom panel.}
\label{fig:sppol}
\end{figure}

In ISPs the peak of the low-energy component ranges from the near-IR through the UV bands. Thus, the 2--8\,keV {\it IXPE} band may include substantial emission from the falling high-frequency tail of the leptonic synchrotron (Sy) spectrum emitted by the most efficiently accelerated electrons and positrons, by a related higher-frequency peak hadronic component (synchrotron emission from protons), or by the flatter-spectrum synchrotron self-Compton (SSC) leptonic emission. Indeed, in leptonic models the {\it IXPE} band may lie in the U-shaped transition region from synchrotron to SSC emission \citep[e.g.,][\cref{fig:sppol}]{peirson_testing_2022}. Since the polarization is expected to differ between Sy and SSC, {\it IXPE} ISP observations can probe both radiation processes, and possibly the jet's composition \citep{peirson_testing_2022, zhang_x-ray_2013}. The latter is of particular interest, as blazars have been proposed as candidate sources of TeV neutrinos and ultra-high-energy cosmic-rays (UHECR), which would require a significant hadronic component in some blazar jets \citep{gao_modelling_2019}. The possible 3$\sigma$ association of ISP blazar TXS~0506+056 with the neutrino IceCube-170922A event motivates this connection \citep{kintscher_icecube_2017, icecube_collaboration_neutrino_2018}. However, the peculiar $\gamma$-ray behavior of TXS~0506+056 and the $\sim$40\% probability of an atmospheric origin of the neutrino challenges the association. Current blazar models that include neutrino emission assume either lepto-hadronic X-ray emission (e.g., \citealp{cerruti_leptohadronic_2019}) or subdominant hadronic components where the proton emission only dominates the SED in the transition region where the leptonic component is at a minimum level (e.g., \citealp{gao_modelling_2019}). These factors highlight the importance of distinguishing between leptonic and hadronic emission in an ISP blazar when it comes to X-ray polarization measurements with {\it IXPE}.

BL Lac is among the few LSP or ISP blazars detected at $\geq0.1$ TeV (very high-energy, VHE) $\gamma$-ray energies, and is the 14th brightest AGN at GeV energies in the {\it Fermi} 4LAC catalog \citep{ajello_fourth_2020}. It is a rapidly variable VHE source on timescales as short as $<1$ hour \citep{albert_discovery_2007, arlen_rapid_2013}. Moreover, BL Lac has been the focus of a large number of multi-wavelength and polarization studies \cite[e.g.,][]{blinov_robopol:_2015, weaver_multiwavelength_2020, raiteri_awakening_2013, casadio_jet_2021}. Because of this blazar's strong variability, the interpretation of its SED requires simultaneous observations.

In this paper, we report the first $> 99\%$ confidence detection of X-ray polarization in BL Lac, with redshift $z = 0.0686$ \citep{vermeulen_when_1995} and synchrotron peak frequency $\nu_{\rm syn}=1.98\times10^{14}~{\rm Hz}$ \citep{chen_curvature_2014}. Although BL Lac is typically classified as an LSP, where the hard secondary SED component dominates the soft X-ray band, its recent outburst both softened and brightened its X-ray spectrum. The soft X-ray flux and photon index are highly variable, but typically $<2 \times 10^{-11}$ erg cm$^2$s$^{-1}$ and $< 2$, respectively \citep[e.g.,][]{wehrle_erratic_2016,giommi_x-ray_2021, sahakyan_13-yr-long_2022, middei_first_2022}. However, during the recent flare the \textit{Swift} XRT finds a flux $ F_{2-8 {\rm keV}} = (2.77\pm 0.21) \times 10^{-11}$ erg cm$^2$s$^{-1}$, photon index $2.10 \pm 0.09$) on 2022 November 12, which suggests a significant contribution from the synchrotron component.

In \S\ref{sec:analysis} we describe the X-ray, optical, and radio polarization observations and data reduction. We discuss our findings in the context of multiwavelength observations in \S\ref{sec:disc_conc}. Further analysis details can be found in the appendix \S\ref{sec:app_pol}, \S\ref{sec:app_multi}.

\section{Data analysis} 
\label{sec:analysis}

BL Lac was observed by {\it IXPE} with an exposure time of 287~ksec on 2022 November 27--30 (MJD 59910.58 -- 59913.90). {\it IXPE}, launched on 2021 December 9, is a joint mission of NASA and the Italian Space Agency (Agenzia Spaziale Italiana, ASI). A description of the instrument is given by \citet{weisskopf_imaging_2022}. At the $\sim30''$ angular resolution of {\it IXPE}, BL Lac is a point-like source. Quasi-simultaneously, BL Lac was observed in the hard and soft X-ray bands with \textit{NuSTAR} (MJD 59911.87 -- 59912.31), \textit{XMM-Newton} (MJD 59910.27 -- 59910.39) and \textit{Swift} XRT (MJD 59910.16, 59911.89, 59912.22, 59913.43), in linear polarization at optical BVRI bands with Calar Alto, the Nordic Optical Telescope (NOT), the 1.8m Perkins Telescope of Boston University, and the Sierra Nevada Observatory, and at millimeter wavelengths by the Institut de Radioastronomie Millim\'{e}trique 30-m Telescope (IRAM-30m) and SubMillimeter Array (SMA). Additional low-frequency radio observations were performed with the Effelsberg 100m telescope (4.85-10.45~GHz) and KVN (22-123~GHz). Appendices \ref{sec:app_xray}, \ref{sec:app_multi} detail the data reduction of these multi-wavelength observations. The {\it IXPE} data were calibrated and reduced following standard procedures within the {\it ixpeobssim} pipeline\footnote{\url{https://agenda.infn.it/event/15643/contributions/30820/attachments/21780/24810/ixpeobssim.pdf}} \citep{baldini_ixpeobssim_2022}. 

We begin by measuring the model-independent, constant polarization parameters for the full IXPE observation using \textit{ixpeobssim}'s PCUBE algorithm, finding 99\% upper limits on the polarization fraction of 14.3\% (2--8 keV) and 19.8\% (2--4 keV).

To characterize the contribution of the two emission components in the {\it IXPE} 2--8 keV band and identify a fiducial spectral model, we first attempt to fit the broad-band X-ray spectrum averaged over the full observation, including \textit{XMM-Newton} 0.3--10 keV and \textit{NuSTAR} 3--80 keV data. A simple absorbed power-law model is inadequate, with $\chi^2/{\rm dof} = 1898/1087$. We find the sum of two absorbed power-laws best reproduces the observations, as measured by $\chi^2$ per degree of freedom, \cref{tab:xspec}. {\it XMM-Newton} measurements below 1~keV require an additional soft spectral component, probably from a hot diffuse plasma, as noted for BL Lac by \citet{middei_first_2022}; we thus add an unpolarized apec component in XSPEC. The temperature and normalization of this apec component are free parameters in the fit. The best-fit absorption column density, $N_H = 3.1\pm0.12\times10^{21}$ cm$^{-2}$, is consistent with values from previous studies \citep[e.g., $N_H = 2.8\pm0.05\times10^{21}$ cm$^{-2}$;][]{weaver_multiwavelength_2020}. \Cref{fig:sppol} shows the results of the double power-law fit, including the individual power-law and apec components. We note that the soft power-law component comprises $36.7\%$ (2--4\,keV) or $26.1\%$(2--8\,keV) of the total flux.

By extending to a spectro-polarimetric fit and restricting the softer power-law to have constant linear polarization, while assuming that the harder power-law is unpolarized, we find $\Pi_X = 27.6\% \pm 11.6$\% (with $\Pi_X < 57.6$\% at 99\% confidence), $\psi_X = -34.5^{\circ} \pm 11.6^{\circ}$. We do not consider this a significant detection, since it does not meet our 99\% confidence of non-zero polarization requirement. 
We discuss the rationale for assuming an unpolarized high-energy component in \textsection\ref{sec:disc_conc}. Allowing both power-laws to have independent, constant polarization results in high, nearly orthogonal values, which cancel the net polarization over most of the spectrum; these values are poorly constrained owing to the poor statistics.

\begin{table}[t]
\centering
\caption{Spectro-Polarimetric XSPEC model fit to joint {\it NuSTAR}, 
{\it XMM-Newton}, and {\it IXPE} spectra, \cref{fig:sppol}.}
\begin{tabular}{lcc}
\toprule
Model Component & $+2$ Power-laws \\
\midrule
$\chi^2 / $dof & $1102/1085$\\
$N_H$ [$10^{21} {\rm cm}^2$] & $3.10 \pm 0.12$\\
$kT$ [keV] & $0.345 \pm 0.025$\\
$\Gamma_1$ & $3.28 \pm 0.100$\\
$\Gamma_2$ & $1.62 \pm  0.027$\\
$\Pi_1$ [\%] & $27.6 \pm 11.6$\\
$\psi_1$ [$^{\circ}$] & $-34.5 \pm 11.6$ \\
\bottomrule
\end{tabular}
\tablecomments{$\Gamma$ -- photon index, $\Pi$ -- polarization fraction, $\psi$ -- EVPA. The polarization fraction of the second power-law component is fixed to zero.}
\label{tab:xspec}
\end{table}

\subsection{Time variability}

\begin{figure}[t]
\includegraphics[width=0.4475\textwidth]{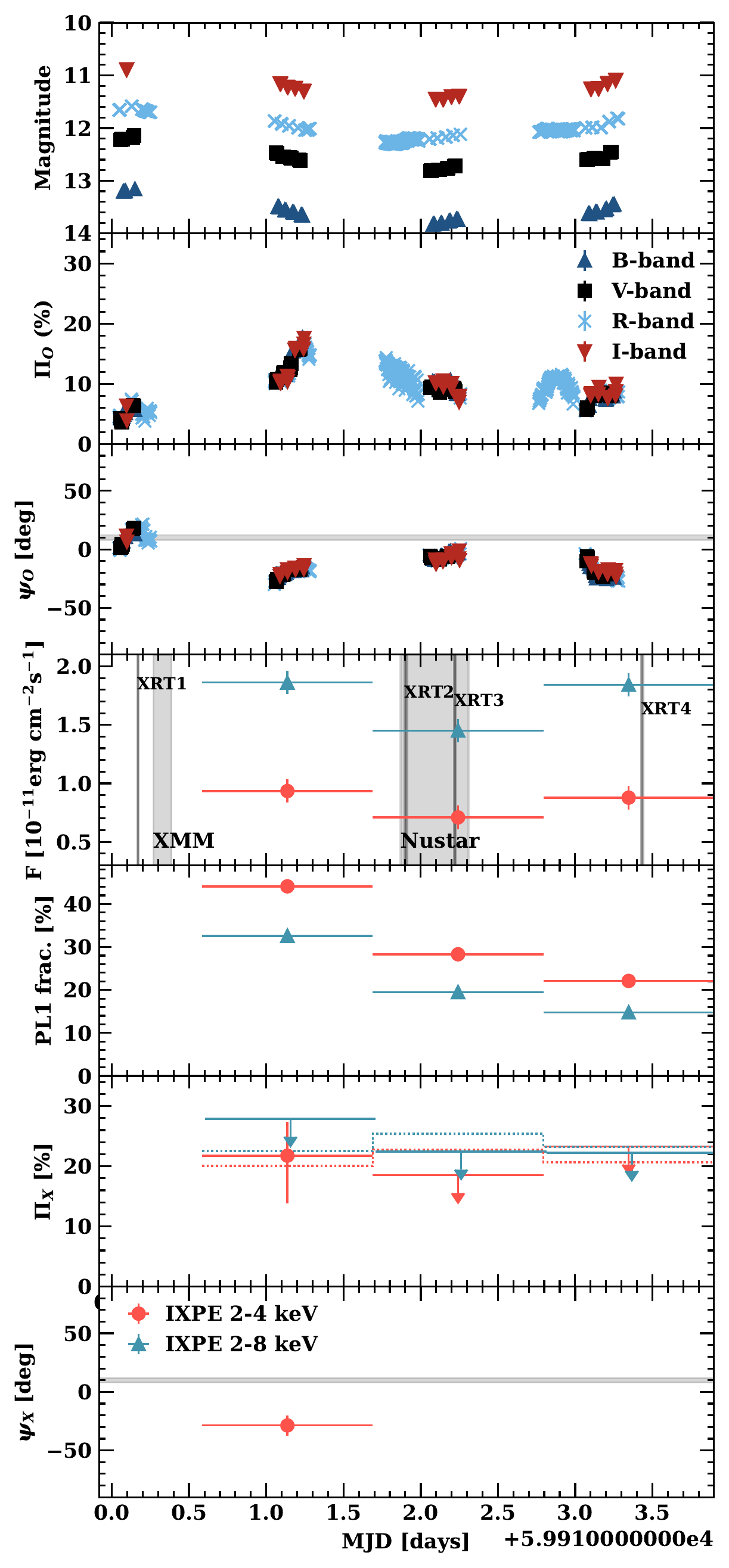}
 \caption{Light curves, polarization fractions, and EVPAs in the optical (top three panels) and X-ray (bottom three panels) bands. X-ray soft power flux contribution fraction is shown in the middle panel. In the X-ray panels, the data are split into two energy bins and three equal time bins. Times of observations from satellites other than IXPE are indicated in the flux panel. Errors are $68.3\%$ (1$\sigma$) confidence intervals and upper limits are $99\%$ confidence. Shaded horizontal regions in EVPA panels represent the jet axis projection on the plane of the sky \citep{weaver_kinematics_2022}. Dotted lines in the X-ray polarization-fraction panel represent the MDP99 level for each bin individually. For X-ray polarization measurements, only the first time bin at 2--4 keV exceeds MDP99.}
\label{fig:time}
\end{figure}

The null result is based on the assumption of constant polarization with time of both power-law spectral components. However, the source is clearly variable during the present {\it IXPE} epoch, which may affect the derived polarizations. To assess this, we split the observation into three equal time bins based on the {\it IXPE} count spectrum and optical observations. We also analyze the data over two energy bins, measuring separately the low- (2--4\,keV) and full-energy (2--8\,keV) ranges.
To check whether the polarization degree and electric vector position angles (EVPA) vary with energy, supporting a multi-component interpretation, we make model-independent measurements of constant polarization in each time and energy bin, \cref{fig:time}.

{\it XMM-Newton} and {\it NuSTAR} observations straddle the first time bin, and three short $\sim 1$\,ksec \textit{Swift} XRT exposures are available, two within the second time bin and the other in the final time bin. To assess the contribution of the individual power-law components to each time bin, we fit the absorbed sum of a double power-law model (\cref{tab:xspec}) separately to each appropriate set of {\it IXPE} plus {\it Swift} XRT time-binned spectra, fixing the photon indices $\Gamma_1$ and $\Gamma_2$ to their {\it NuSTAR} plus {\it XMM-Newton} fitted values. This allows the power-law normalizations to vary in time while their slopes remain fixed. Estimated fractions of the soft power-law (synchrotron) contribution for each time and energy bin are displayed in \cref{fig:time}. While the similar variation of the fluxes in the two energy bins (\cref{fig:time}) shows that both power-laws vary together (as expected for synchrotron and Compton components), we find that the soft-component contribution is largest in the first time bin, as might  be expected from observations at the high-energy tail of a cooling synchrotron flare. 

\begin{figure}[t]
  \centering
  \includegraphics[width=1\linewidth]{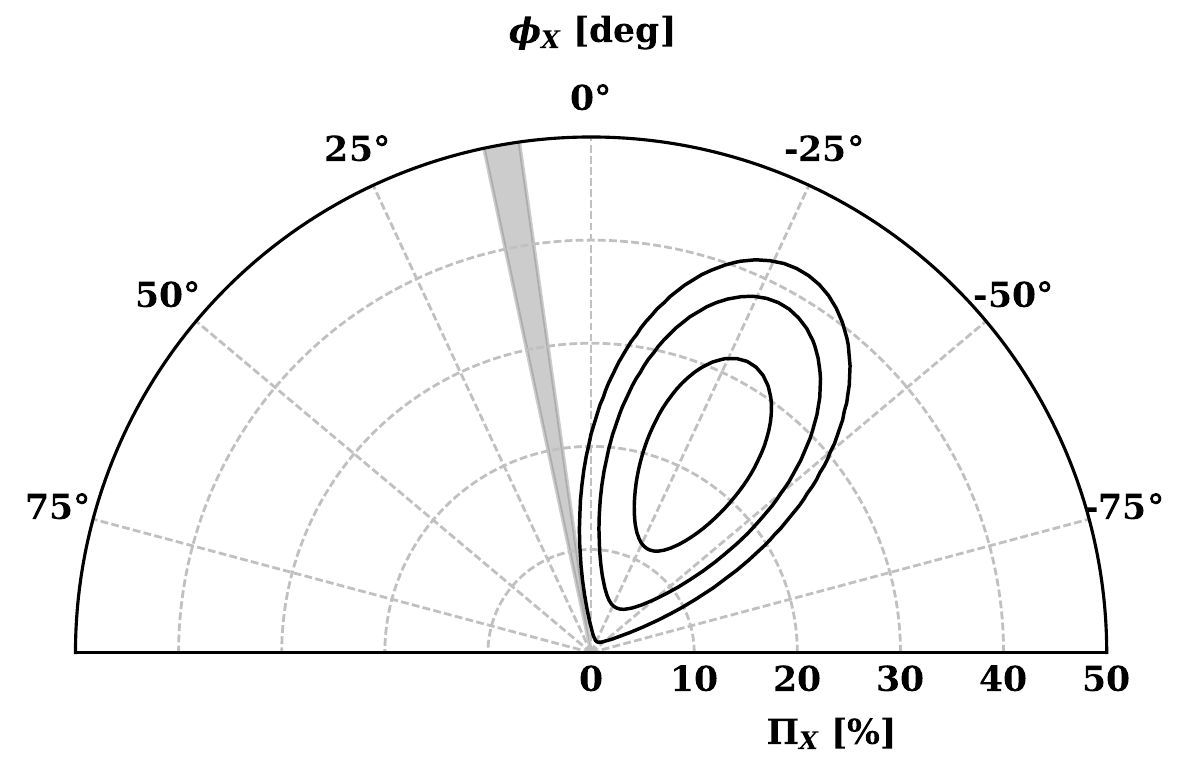}
  \caption{Polarization fraction and EVPA confidence levels (68\%, 95\%, 99\%) for the first time bin over the  2--4 keV energy range, \cref{fig:time}.
  Gray shaded region represents the VLBI-determined jet axis projection on the plane of the sky \citep{weaver_kinematics_2022}.}
  \label{fig:radar}
\end{figure}

Interestingly, in the low energy channel within the first time bin, where our spectral fits indicate that the soft component contributes the maximum flux, we detect linear polarization exceeding minimum detectable polarization at 99\% confidence (MDP99), \cref{fig:radar}. When considered as a single measurement, we find $99.3\%$ confidence in non-zero polarization. Considering all three binned 2--4 keV polarization measurements jointly we find a $98.4\%$ confidence in non-zero polarization.
As noted previously, we do not exceed the 99\% threshold over the full time interval.

We have attempted full spectro-polarimetric double power-law fits to each time bin as in \cref{fig:sppol}, but none produced a high-significance detection for the soft power-law component. Instead, we have used these fits to estimate the lower-energy power-law (synchrotron) contribution fractions in \cref{fig:time}. The results for all such fits are detailed in Appendix \ref{sec:app_pol}. 

\section{Discussion \& Conclusions}\label{sec:disc_conc}

In order to fit our \textit{XMM-Newton} and \textit{NuSTAR} spectra, (Fig.~\ref{fig:sppol}), we find that a two-component spectrum is required for BL Lac. A power-law fit with low-energy photon index $\Gamma_1 = 3.28$ and high-energy value $\Gamma_2 = 1.62$, with equal flux at 2\,keV, is statistically preferred over a simple power-law. Our polarization results thus probe the cross-over region. 

Pure leptonic models predict a significant decrement in polarization of the hard (Compton) component relative to the optical synchrotron value, 
with $\Pi_{SSC}/\Pi_{Sy} \approx 0.3$ \citep{krawczynski_polarization_2012, peirson_polarization_2019}. When the high-energy tail of the synchrotron emission reaches into the {\it IXPE} band, we might expect the soft-component polarization even to exceed the optical value if the X-ray emission arises from a region of more highly-ordered
magnetic field.

One possibility is that the X-ray synchrotron emission occurs mainly in the acceleration and collimation zone of the jet, where the magnetic field is expected to have a well-ordered helical geometry \citep[e.g.,][]{vlahakis_magnetic_2004}, as inferred previously in BL Lac \citep{marscher_inner_2008}. Alternatively, the emission could take place in a region of the jet with a turbulent magnetic field, with acceleration of particles occurring only within a small volume, for example by magnetic reconnection. Since X-ray synchrotron radiation requires extremely high-energy electrons that are subject to strong radiative energy losses, the X-ray emission would be confined to locations close to the site of particle acceleration, while lower-frequency emission from lower-energy electrons, which can travel farther from the acceleration site before their fractional energy loss becomes high, would occur over a larger volume. In a turbulent magnetic field, the net field is more ordered (but randomly oriented) over smaller volumes, hence the polarization is higher \citep[see, e.g.,][ for discussions and estimates]{marscher_turbulent_2014,peirson_polarization_2018}. The random value of $\psi$ agrees with our finding that the observed EVPA does not appear related to the direction of the jet.

If the turbulent plasma encounters a shock, the magnetic field becomes partially aligned with the shock front \citep[e.g.,][]{hughes_polarized_1985,marscher_turbulent_2014,tavecchio_probing_2018}, at which X-ray emitting particles can be accelerated. By the same argument as above, the synchrotron X-ray polarization should then be higher than at longer wavelengths, but instead of a random orientation of the EVPA, $\psi$ should be oriented along the jet direction, contrary to the observations of BL Lac presented here. 

During a synchrotron X-ray flare, the more highly polarized synchrotron component contributes more to the X-ray spectrum at lower X-ray energies than does the flatter-spectrum Compton component. As this steep-spectrum component fades across the {\it IXPE} band, the lower Compton polarization should dominate at higher energies. Although the EVPAs measured at the synchrotron peak and SSC peak frequencies are expected to be correlated \citep{peirson_polarization_2019}, turbulence and energy-dependent emission volumes can cause differences.

To best constrain the soft component polarization we assume that the high energy power-law component is negligibly polarized, as the logical approximation of the expected polarization decrease noted above. The polarization $\Pi_X = 27.6 \% \pm 11.6\%$ measured over all {\it IXPE} times and energies, while a $>2\sigma$ detection, does not reach the threshold for high significance. However, noting that the soft component varies in strength, we find that a time and energy bin analysis allows a significant $\Pi_X = 21.7^{+5.6}_{-7.9}\%$ detection at low energies in the first time bin, when the spectral analysis indicates that the soft component accounts for the largest fraction of the total flux. Note that this polarization is substantially higher than the simultaneous optical polarization $\Pi_O = 13.1\% \pm 2.4\%$ for the same time bin. Both the optical and X-ray EVPA are at $\sim 40^{\circ}$ to the projection of BL Lac's jet axis on the plane of the sky at 43GHz \citep{weaver_kinematics_2022}.  

Even the first time/energy bin is diluted by the high energy component. Under the assumption of relatively constant photon indices at the {\it XMM-Newton} and {\it NuSTAR} values, the high-energy component still comprises $56\%$ of the flux in this 2--4~keV time bin (\cref{fig:time}). If the component is unpolarized, correction for this flux implies $ \Pi_{Sy} = 21.7\% / (1 - 0.56) \approx 49\%$ for the soft component; if the 
high-energy component corresponds to Compton scattering, polarized at $\sim 0.3\times$ the synchrotron value with approximately the same EVPA, the polarization of the low energy component is $ \Pi_{Sy} = 21.7\% / ((1 - 0.56) + 0.3 \times 0.56) \approx 36\%$. These polarization degrees are consistent with the upper limit provided by the double power-law spectral fit in \cref{tab:time_xspec} of the appendix. These large values dramatically exceed the optical polarization, arguing that we are probing the high-frequency end of the synchrotron component, which fades during our observation window.

Two previous {\it IXPE} observations of BL Lac \citep{middei_x-ray_2023}, which occurred during a low state, where the X-ray spectrum was apparently dominated by SSC, produced upper limits on $\Pi_X$ that were significantly below the simultaneous optical polarization. However, in the data from the third epoch presented here, even the total (time and energy) averaged {\it IXPE} polarization signal is comparable to the optical values. After correction for hard-component contamination, the soft component substantially exceeds the optical polarization degree. 

Other scenarios have different polarization signatures. In simple single-zone lepto-hadronic models, the polarization should be similar across the transition region (fig.~\ref{fig:sppol}). Synchrotron emission from primary leptons producing the low-energy hump yields to higher-energy synchrotron radiation from protons and secondary leptons produced in $p\gamma$ cascades \citep{zhang_x-ray_2013}, all in the same magnetic field environment. 
As noted above, in more realistic jet models, turbulent magnetic fields, differential cooling times \citep{marscher_probing_2010} and relativistic boosting \citep{peirson_polarization_2019} cause a significant increase in polarization and changes in EVPA as one approaches the quasi-exponential cutoff of the low-energy peak, where only the highest energy electrons from the most efficient particle acceleration zones contribute synchrotron emission.
If a hadronic emission component becomes dominant above this energy, it should be averaged over the lower-energy particle population typical of particle acceleration in the jet, and thus should display polarization similar to that of the primary (optical) synchrotron peak. In models including hadronic synchrotron emission, we expect the polarization fraction at the upper end of the transition region to meet or exceed the optical value.

Our past measurements of $\Pi_X < \Pi_{O}$ \citep{middei_x-ray_2023} and the persistent low polarization at the upper end of the {\it IXPE} band therefore argues against a lepto-hadronic emission model. Furthermore, the detection of total polarization at low energies, with $\Pi_{\rm 2-4 keV} \gg \Pi_{O}$, indicates that we are sampling the upper cut-off of the synchrotron spectrum. This is even more striking after correction for dilution by the hard, weakly polarized component. The decreasing relative flux of the soft component and the resulting loss of detection of polarization are consistent with the tail end of a jet flare, as suggested by the $\gamma$-ray and X-ray light curves.

While these results provide tentative evidence for a fully leptonic emission model in BL Lac at our {\it IXPE} observation epoch, variability allows significant hadronic emission at other times. Blazars are particularly variable in polarization \citep{blinov_robopol:_2018}, and broadband or long-exposure measurements can make polarization results difficult to interpret properly \citep{kiehlmann_time-dependent_2021, peirson_testing_2022}. Thus, repeated polarization measurements of blazars, with attendant contemporaneous multiwavelength intensity and polarization observations, are needed to fully explore blazar emission. Most interesting would be measurements during flares associated with neutrino events. With the plausible neutrino connection, and the ability to probe the critical Sy/SSC transition regions, BL Lac and other ISPs will be prime targets for such studies.
\bigskip

\facilities{Calar Alto, Effelsberg-100m, IRAM-30m, {\it IXPE}, Kanata, KVN, Nordic Optical Telescope, {\it NuSTAR}, Perkins, SMA, {\it XMM-Newton}, {\it Swift}}

\section{acknowledgments}
The Imaging X-ray Polarimetry Explorer ({\it IXPE}) is a joint US and Italian mission.  The US contribution is supported by the National Aeronautics and Space Administration (NASA) and led and managed by its Marshall Space Flight Center (MSFC), with industry partner Ball Aerospace (contract NNM15AA18C).  The Italian contribution is supported by the Italian Space Agency (Agenzia Spaziale Italiana, ASI) through contract ASI-OHBI-2017-12-I.0, agreements ASI-INAF-2017-12-H0 and ASI-INFN-2017.13-H0, and its Space Science Data Center (SSDC) with agreements ASI-INAF-2022-14-HH.0 and ASI-INFN 2021-43-HH.0, and by the Istituto Nazionale di Astrofisica (INAF) and the Istituto Nazionale di Fisica Nucleare (INFN) in Italy.  This research used data products provided by the {\it IXPE} Team (MSFC, SSDC, INAF, and INFN) and distributed with additional software tools by the High-Energy Astrophysics Science Archive Research Center (HEASARC), at NASA Goddard Space Flight Center (GSFC).  Funding for this work was provided in part by contract 80MSFC17C0012 from the MSFC to MIT in support of the {\it IXPE} project.  Support for this work was provided in part by the NASA through the Smithsonian Astrophysical Observatory (SAO)
contract SV3-73016 to MIT for support of the {\it Chandra} X-Ray Center (CXC), which is operated by SAO for and on behalf of NASA under contract NAS8-03060. The IAA-CSIC co-authors acknowledge financial support from the Spanish "Ministerio de Ciencia e Innovaci\'{o}n" (MCIN/AEI/ 10.13039/501100011033) through the Center of Excellence Severo Ochoa award for the Instituto de Astrof\'{i}isica de Andaluc\'{i}a-CSIC (CEX2021-001131-S), and through grants PID2019-107847RB-C44 and PID2022-139117NB-C44. Some of the data are based on observations collected at the Observatorio de Sierra Nevada, owned and operated by the Instituto de Astrof\'{i}sica de Andaluc\'{i}a (IAA-CSIC). Further data are based on observations collected at the Centro Astron\'{o}mico Hispano-Alem\'{a}n(CAHA), operated jointly by Junta de Andaluc\'{i}a and Consejo Superior de Investigaciones Cient\'{i}ficas (IAA-CSIC). The POLAMI observations were carried out at the IRAM 30m Telescope. IRAM is supported by INSU/CNRS (France), MPG (Germany) and IGN (Spain). The Submillimetre Array is a joint project between the Smithsonian Astrophysical Observatory and the Academia Sinica Institute of Astronomy and Astrophysics and is funded by the Smithsonian Institution and the Academia Sinica. Mauna Kea, the location of the SMA, is a culturally important site for the indigenous Hawaiian people; we are privileged to study the cosmos from its summit. The data in this study include observations made with the Nordic Optical Telescope, owned in collaboration by the University of Turku and Aarhus University, and operated jointly by Aarhus University, the University of Turku and the University of Oslo, representing Denmark, Finland and Norway, the University of Iceland and Stockholm University at the Observatorio del Roque de los Muchachos, La Palma, Spain, of the Instituto de Astrofisica de Canarias. The data presented here were obtained in part with ALFOSC, which is provided by the Instituto de Astrof\'{\i}sica de Andaluc\'{\i}a (IAA) under a joint agreement with the University of Copenhagen and NOT. E.\ L.\ was supported by Academy of Finland projects 317636 and 320045. We acknowledge funding to support our NOT observations from the Finnish Centre for Astronomy with ESO (FINCA), University of Turku, Finland (Academy of Finland grant nr 306531). The research at Boston University was supported in part by National Science Foundation grant AST-2108622, NASA {\it Fermi} Guest Investigator grants 80NSSC21K1917 and 80NSSC22K1571, and NASA {\it Swift} Guest Investigator grant 80NSSC22K0537. This study used observations conducted with the 1.8 m Perkins Telescope Observatory (PTO) in Arizona (USA), which is
owned and operated by Boston University. The above study is based in part on observations obtained 
with {\it XMM-Newton}, an ESA science mission with instruments and contributions directly funded by ESA Member States and NASA. We are grateful to the 
{\it NuSTAR} team for approving our DDT request.  This work was supported under NASA contract No. NNG08FD60C, and made use of data from the {\it NuSTAR} mission, a project led by the California Institute of Technology, managed by the Jet Propulsion Laboratory, and funded by the NASA. This research has made use of the {\it NuSTAR} Data Analysis Software (NuSTARDAS) jointly developed by the ASI Science Data Center (ASDC, Italy) and the California Institute of Technology (USA). This work was supported by JST, the establishment of university fellowships towards the creation of science technology innovation, Grant Number JPMJFS2129. This work was supported by Japan Society for the Promotion of Science (JSPS) KAKENHI Grant Numbers JP21H01137. This work was also partially supported by Optical and Near-Infrared Astronomy Inter-University Cooperation Program from the Ministry of Education, Culture, Sports, Science and Technology (MEXT) of Japan. We are grateful to the observation and operating members of Kanata Telescope. MN acknowledges the support by NASA under award number 80GSFC21M0002. CC acknowledges support by the ERC under the Horizon ERC Grants 2021 programme under grant agreement no. 101040021. SK, S-SL, WYC, S-HK, and H-WJ were supported by the National Research Foundation of Korea (NRF) grant funded by the Korea government (MIST) (2020R1A2C2009003). The KVN is a facility operated by the Korea Astronomy and Space Science Institute. The KVN operations are supported by KREONET (Korea Research Environment Open NETwork) which is managed and operated by KISTI (Korea Institute of Science and Technology Information). Partly based on observations with the 100-m telescope of the MPIfR (Max-Planck-Institut f\"ur Radioastronomie) at Effelsberg. Observations with the 100-m radio telescope at Effelsberg have received funding from the European Union’s Horizon 2020 research and innovation programme under grant agreement No 101004719 (ORP). ALP acknowledges support from NASA FINESST grant 80NSSC19K1407 and the Stanford Data Science Scholars program.

\bibliography{references}
\bibliographystyle{aasjournal}

\newpage

\appendix

\section{X-ray Observations}
\label{sec:app_xray}
Between 2022 November 27--30, BL Lacertae was observed quasi-simultaneously with {\it IXPE}, {\it NuSTAR}, and {\it XMM-Newton}, \cref{tab:obs}. For the {\it IXPE} data, the cleaned event files and the associated science products were obtained using a dedicated pipeline relying on the Ftools software package and adopting the latest calibration data files from {\it IXPE} (CALDB 20211118). The source regions for each of the three detector units (DUs) were then selected via an iterative process aimed at maximizing the signal-to-noise ratio (SNR) over the {\it IXPE} standard energy range of 2-–8 keV. In particular, we used circular regions with radius 50$''$ for all three DUs and annulus regions of size 100--300$''$ for the background spectra in Stokes parameters $I$, $Q$, and $U$. 

To improve the polarimetric sensitivity, we applied a background rejection strategy \citep{xie_study_2021, di_marco_handling_2023}. The method refines the sensitivity by applying energy-dependent cuts on photo-electron tracks from the level 1 data. This method substantially decreases the background event rate, which are triggered by electrons, positrons, muons, or protons either directly impinging upon the detector or created by high energy interactions in the surrounding satellite structures. The cuts employ (1) the number of pixels: the size of the track region of interest (\texttt{NUM\_PIX}); (2) energy fraction: the fraction of the event energy in the track (\texttt{EVT\_FRA}); and (3) border pixels: the number of activated pixels along the boundary of the detector (\texttt{NUM\_TRK}). Here we eliminated the events that do not satisfy the X-ray photon criteria of Di Marco et al.\ (2021, submitted). Using the point source as a monitor, we rejected $\sim30$\%  of the diffuse background, with little impact on the source events. 
For {\it IXPE} spectro-polarization analysis, a constant energy binning of 4$\times 0.05$keV PI channels per bin was used for $Q$ and $U$; we required 30 counts per bin for the intensity spectra.

The {\it XMM-Newton} spectra were produced with standard SAS routines and the latest calibration files. The source spectrum was extracted from a circular (radius= 40$''$) aperture, while the background spectrum was extracted from a blank region on the Epic-pn CCD camera from a circular region of the same size. The resulting spectrum was re-grouped to include at least 30 counts in each bin and to avoid $>3\times$ oversampling of the spectral resolution. The {\it NuSTAR} data were calibrated and cleaned with the {\it NuSTAR} Data Analysis Software (NuSTARDAS7), and employed the nuproducts pipeline using the latest calibration database (v. 20220302). The source spectrum was extracted from a circular radius of 70$''$ aperture; a surrounding 270$''$--370$''$ annulus provided the background.

\begin{table}[h]
\centering
\caption{Major quasi-simultaneous X-ray observations related to the 2022 November 27 {\it IXPE} pointing of BL Lac.}
\begin{tabular}{lcccc}
\toprule
Observatory & Start Time & MJD range & ObsID & Exposure [ksec] \\
\midrule
{\it NuSTAR} & 2022-11-28 20:51:09 & 59911.87 -- 59912.31 & 90801633002 & 38.1 \\
{\it XMM-Newton} & 2022-11-27 06:26:53 & 59910.27 -- 59910.39  & 0902111801 & 10.1 \\
{\it IXPE} & 2022-11-27 14:01:15 & 59910.58 -- 59913.90 & 02005901 & 286.4  \\
\bottomrule
\end{tabular}
\label{tab:obs}
\end{table}

{\it Swift}-XRT exposures were obtained in the context of a monitoring campaign tracking the BL Lac flux level before, during, and after the {\it IXPE} pointing. Scientific products from the {\it Swift}-XRT exposures were derived by using the facilities provided by the Space Science Data Center (SSDC8) of the Italian Space Agency (ASI). In particular, the source spectra were extracted from a source-centered $47''$ radius aperture, with a $120''$ -- $150''$ concentric annulus providing a background spectrum. The events were grouped to include at least 25 counts in each spectral bin. We modeled each of the four XRT spectra as a simple power-law with Galactic photo-electric absorption. We report the 2–8 keV fluxes and the inferred photon indices in \cref{tab:xrt}. The softest spectrum corresponds to the first {\it IXPE} time bin, \cref{fig:time}, for which the highest polarization is measured. As discussed in \S \ref{sec:disc_conc}, softer spectra represent a higher synchrotron fraction. 

\begin{table}[t]
\centering
\caption{{\it Swift}-XRT exposures in the vicinity of the {\it IXPE} pointing of BL Lac. \Cref{fig:bllc} displays the exposures superimposed on the {\it IXPE} light curve.}
\begin{tabular}{lccccc}
\toprule
Start Time & MJD & ObsID & Exposure & Flux (2-8 keV) & $\Gamma$ \\
 & & & [s] & [$10^{-11}$erg cm$^{-2}$s$^{-1}$] & \\
\midrule
2022-11-27T03:56:36 & 59910.16 & 00096990016 & 834 & 0.827 & 2.24 \\
2022-11-28T21:30:17 & 59911.89 & 00089562001 & 1474 & 1.57 & 1.94 \\
2022-11-29T05:13:35 & 59912.21 & 00096990017 & 826 & 2.04 & 2.13 \\
2022-11-30T10:19:01 &  59913.42 & 00096990018 & 895 & 2.22 & 1.78 \\
2022-12-01T02:07:36 &  59914.09 & 00096990019 & 849 & 1.97 & 1.87 \\
\bottomrule
\end{tabular}
\tablecomments{To measure flux and $\Gamma$, single absorbed power-laws were fit to the full XRT 0.3--10 keV range}
\label{tab:xrt}
\end{table}

\begin{figure}[t]
\centering
\includegraphics[width=0.5\textwidth]{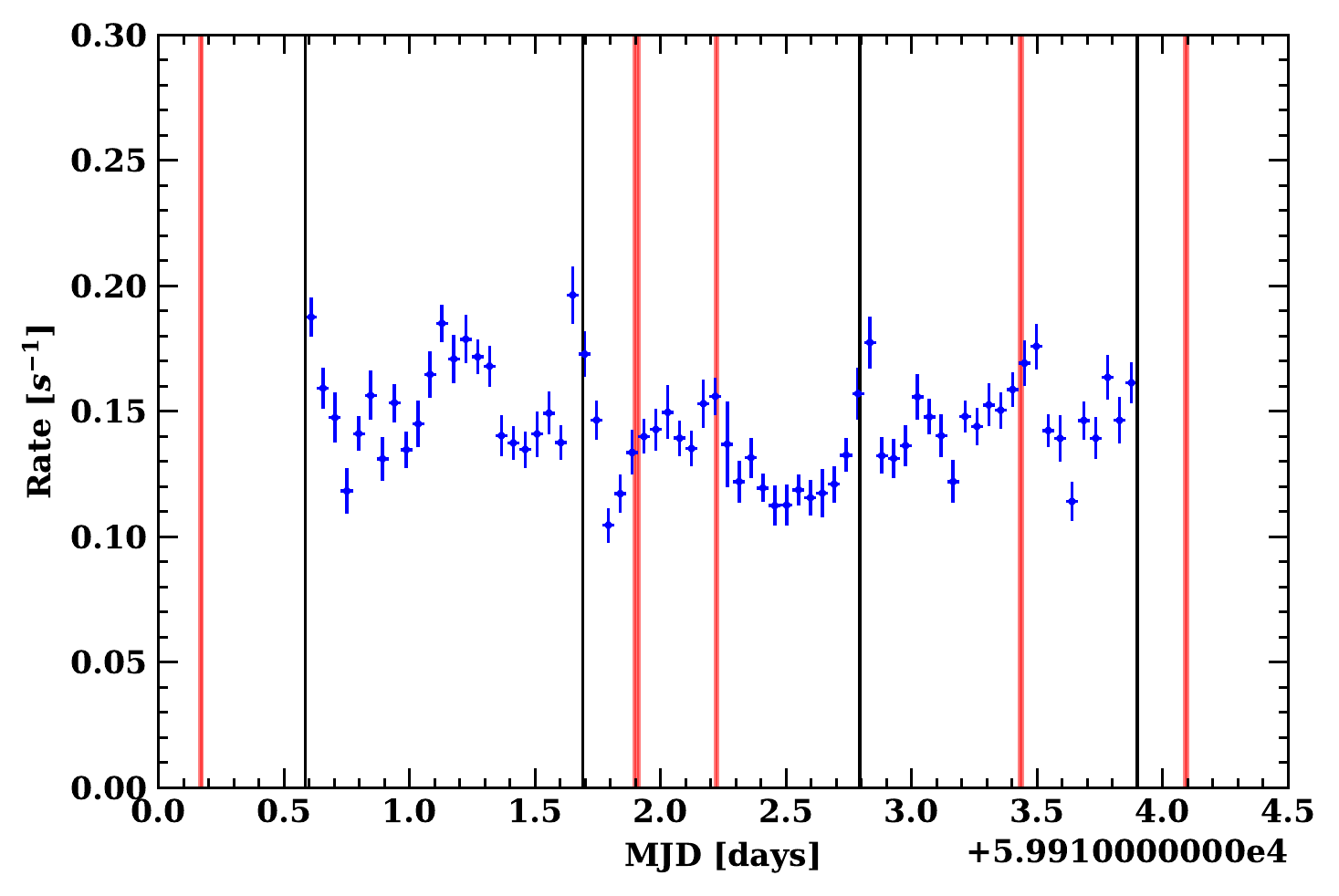}
 \caption{{\it IXPE} light curve. Black lines are three equal time bin separators; red shaded regions indicate XRT exposures.}
\label{fig:bllc}
\end{figure}

\section{Time-resolved spectro-polarimetric analysis.}
\label{sec:app_pol}

\begin{figure}[h]
\centering
\includegraphics[width=0.5\textwidth]{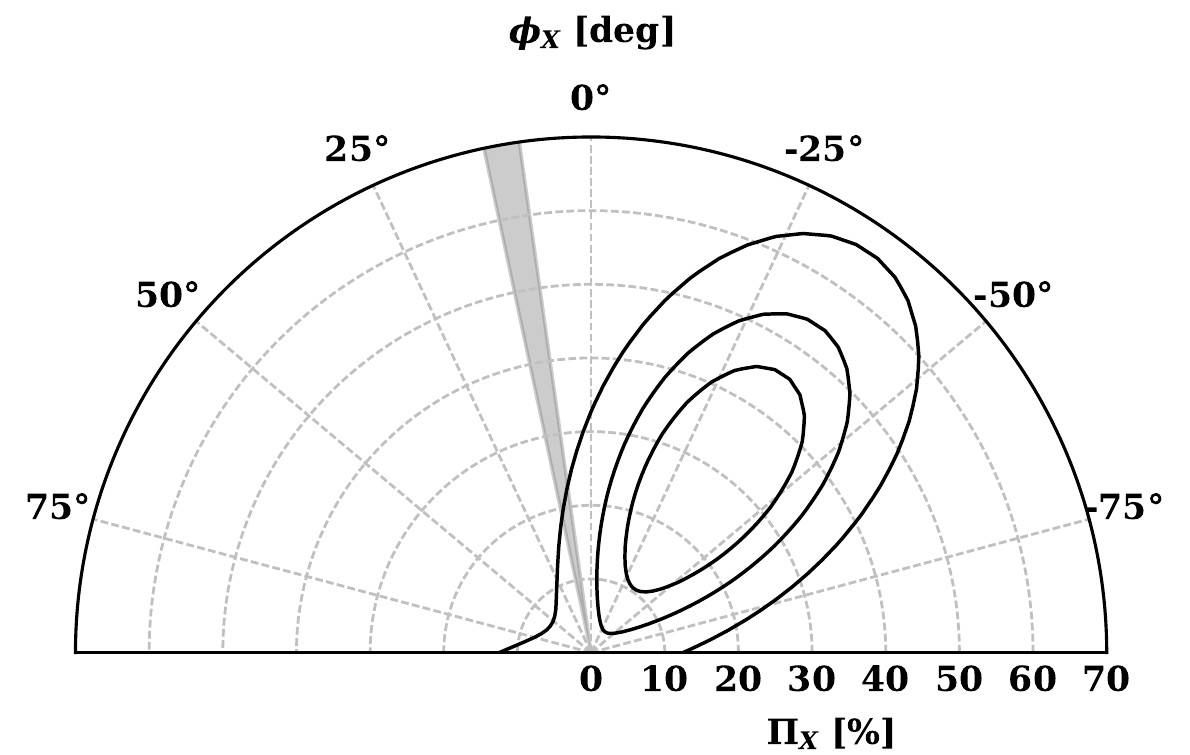}
 \caption{Polarization fraction and EVPA confidence levels (68\%, 95\%, 99\%) for the low-energy power-law in \cref{fig:sppol}. Gray shaded region represents the VLBI-determined jet axis projection on the plane of the sky
 \citep{weaver_kinematics_2022}.}
\label{fig:full_radar}
\end{figure}

We use XSPEC to perform spectro-polarimetric analyses on the entire {\it IXPE} observation, including {\it NuSTAR} and {\it XMM-Newton} spectra, and each of the three {\it IXPE} time bins defined in the main text, \cref{fig:time}. Analysis of the entire observation is discussed in the main text (\cref{fig:sppol}, \cref{tab:xspec}), where we find that an absorbed sum of two power-laws with an apec component,
\begin{verbatim}
    constant * tbAbs * (polconst * powerlaw + polconst * (apec + powerlaw))
\end{verbatim}
is preferred according to XSPEC model fitting. Here we fix polarization degree of both the high-energy power-law (PL) and apec to zero. \Cref{fig:full_radar} displays a contour plot with the polarization measurement of the low-energy PL for the full {\it IXPE} observation fit, as displayed in \cref{fig:sppol}. We note that fixing the polarization degree of the high-energy power-law (PL) to $0.3\times$ the synchrotron, as expected in a leptonic model, does not improve the significance of the fit.

\begin{figure}[ht]
    \centering
    \subfigure[]{\includegraphics[width=0.32\textwidth]{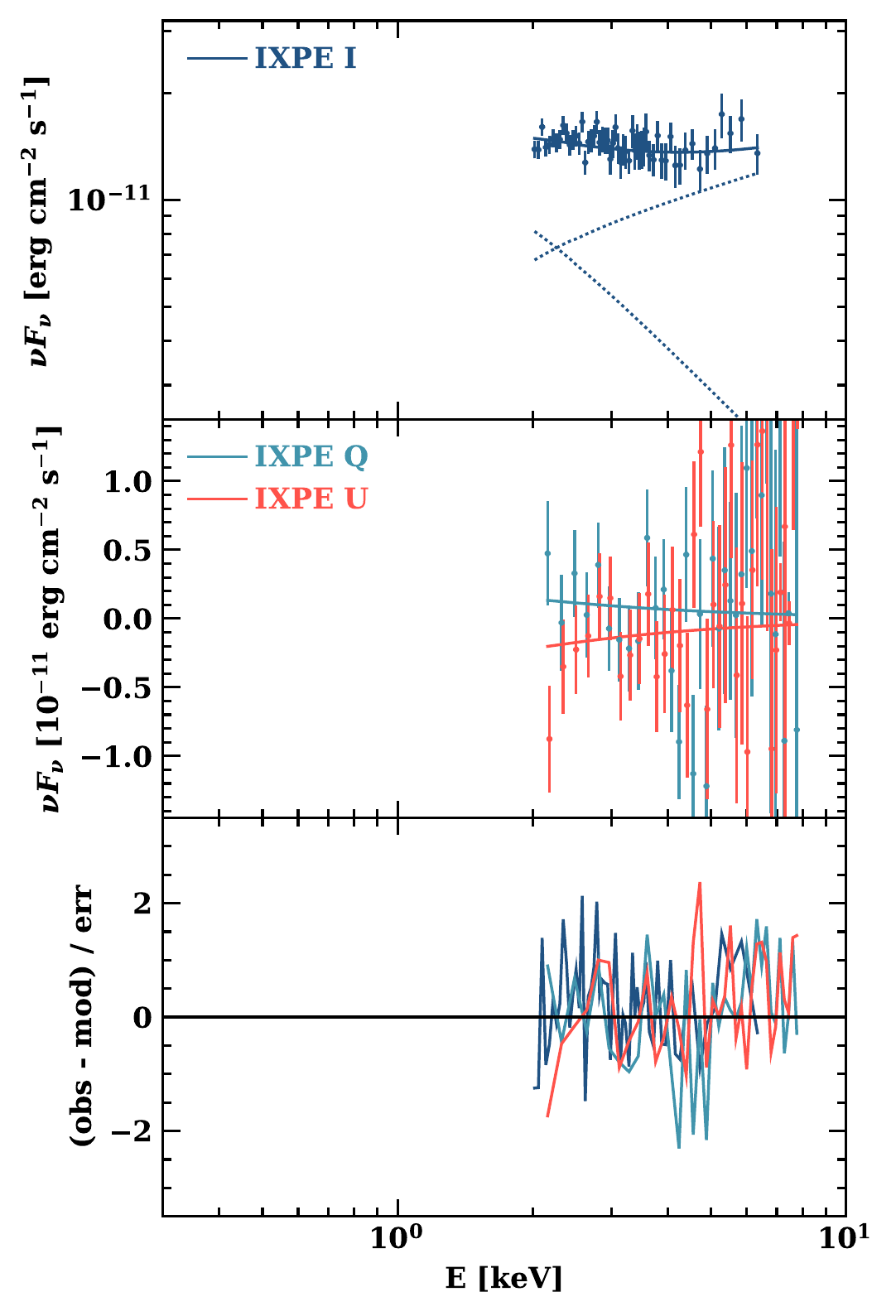}} 
    \subfigure[]{\includegraphics[width=0.32\textwidth]{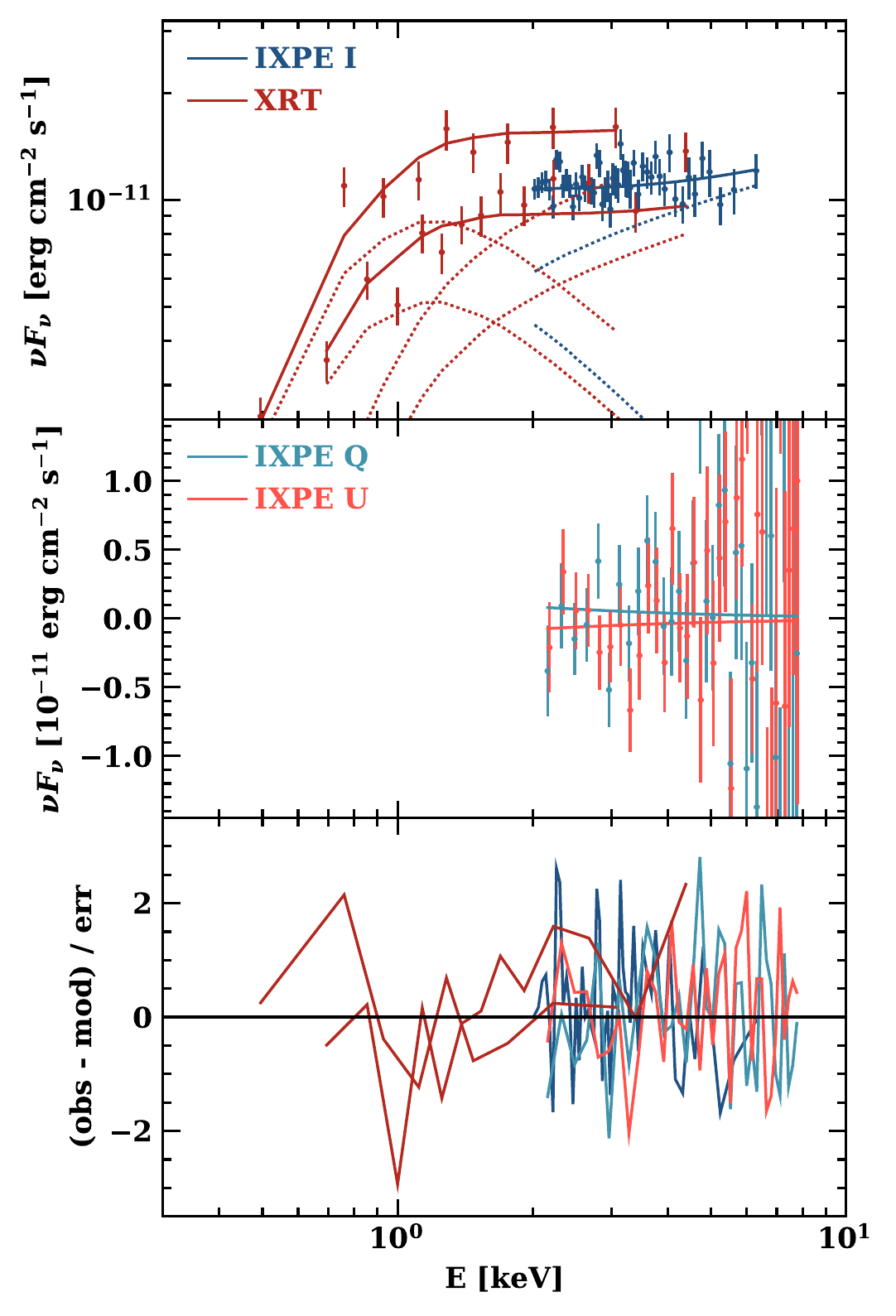}} 
    \subfigure[]{\includegraphics[width=0.32\textwidth]{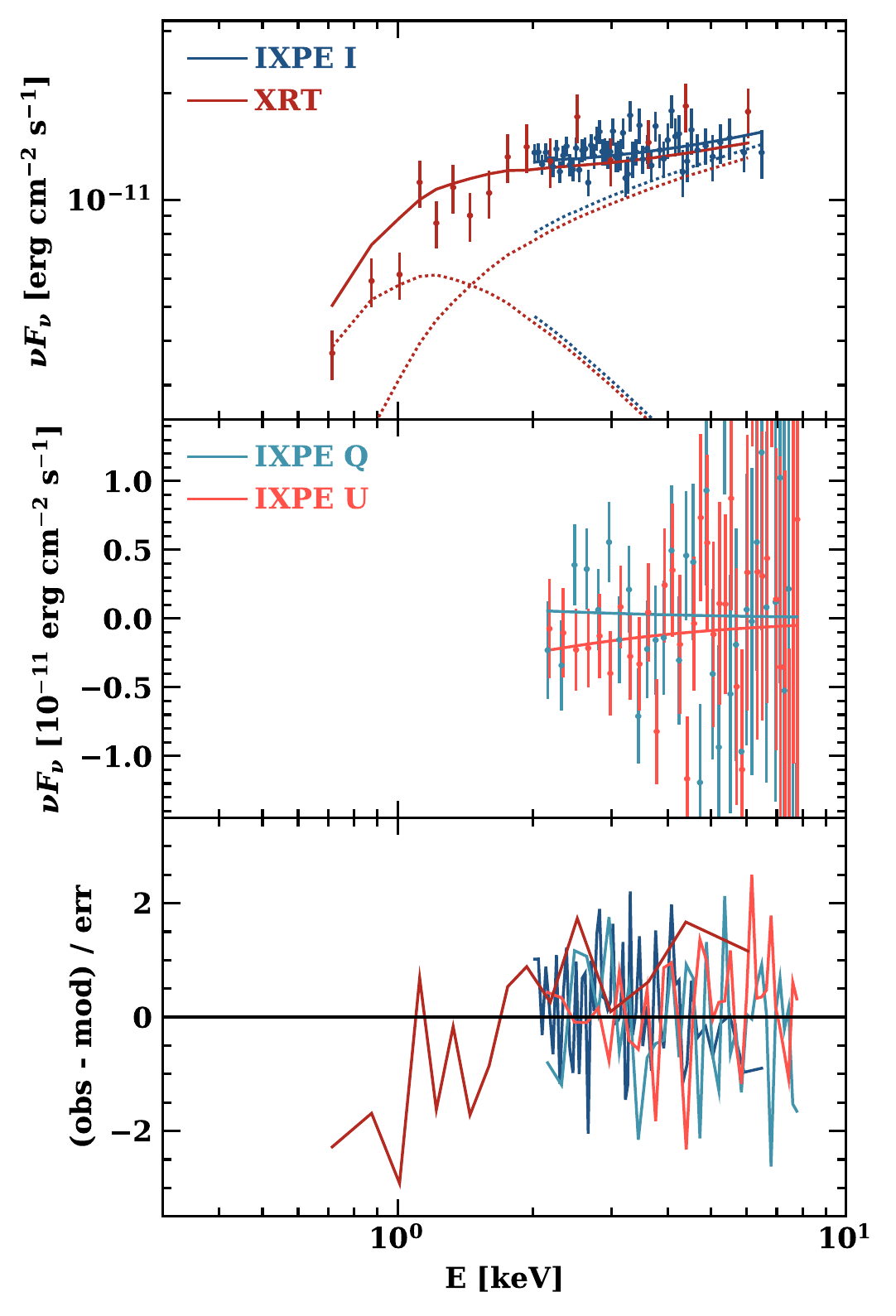}}
    \caption{(a) Absorbed sum of two power-laws with apec fit to first {\it IXPE} time bin. Photon indices and apec temperature are fixed to values in \cref{tab:xspec}. The high-energy power-law polarization is fixed to zero. (b) Second {\it IXPE} time bin with XRT2 and XRT3 observations. (c) Third {\it IXPE} time bin with XRT4 observation.}
    \label{fig:xspec_tfits}
\end{figure}

In the case of the three {\it IXPE} time bins, we use the same model as above, but with photon indices and apec temperature ($\Gamma_1, \Gamma_2, kT$) fixed to the values found in the full observation fit, \cref{tab:xspec}. Hence, only the PL normalization constants and low-energy PL polarization are determined by the fits. We fit this model to the time-binned {\it IXPE} spectra along with the appropriate XRT observation(s); see \cref{fig:time}. The first time bin does not have a simultaneous XRT observation, and the second time bin has two (XRT2, XRT3). \Cref{tab:time_xspec} and \cref{fig:xspec_tfits} give the results of these fits. We are able to determine the relative flux contribution of the low-energy PL in \cref{fig:time}, which correspond to the PL normalization constants at 1 keV.

\begin{table}[h]
\centering
\caption{Sum of two power-law spectro-polarimetric XSPEC model fits to time binned {\it IXPE} and {\it Swift}-XRT spectra, \cref{fig:time}. Photon indices $\Gamma_1, \Gamma_2$, apec temperature $kT$, and $N_H$ are fixed to their full-spectrum fit values; see \cref{tab:xspec}. High-energy PL polarization is fixed to zero.}
\begin{tabular}{lccc}
\toprule
 & \multicolumn{3}{c}{Time bin}\\
\cmidrule{2-4}\\
Model Component & 1 & 2 & 3 \\
\midrule
$\chi^2 / $dof & $393/386$ & $461/382$ & $406/392$\\
PL1 norm & $(4.33 \pm 0.32) \times 10^{-3} $   & $(2.28 \pm  0.23)\times 10^{-3}$ & $(2.44  \pm  0.26)\times 10^{-3}$ \\
PL2 norm & $(1.14  \pm  0.062)\times 10^{-3}$ & $(1.00  \pm  0.49)\times 10^{-3}$  & $(1.31 \pm  0.056)\times 10^{-3}$  \\
$\Pi_1$ [\%] & 
$< 69.5$ & 
$< 89.9$ & 
$< 98.0$ \\
$\psi_1$ [$^{\circ}$] &  
n/a & 
n/a & 
n/a \\
\bottomrule
\end{tabular}
\tablecomments{$\Pi$ -- polarization fraction, $\psi$ -- EVPA. Relative power-law normalizations in each time bin dictate the soft power-law fractions in \cref{fig:time}.}
\label{tab:time_xspec}
\end{table}

\section{Multiwavelength observations}
\label{sec:app_multi}

During the {\it IXPE} observation, BL Lac was contemporaneously observed in polarization by different telescopes at radio, millimeter, and optical wavelengths. Low frequency radio observations (4.85-10.45~GHz) were performed using the Effelsberg 100m telescope through the QUIVER program (Monitoring the Stokes Q, U, {I} and {V} {E}mission of AGN jets in {R}adio) on 2022 December 1 (MJD~59914.94424), and the Korean VLBI Network (KVN, 22-129~GHz) on the 2022-11-30 (MJD~59913.6). Millimeter-wave (mm) observations were performed by the Institut de Radioastronomie Millim\'{e}trique 30-m Telescope (IRAM-30m) on the 2022 November 28 (MJD~59911.5717) at 1.3 mm (228.93 GHz) and 3.5 mm (86.24 GHz), and by the Submillimeter Array (SMA) at 1.3 mm (225.538~GHz) on the 2022 December 1 (MJD~59914.0).

The QUIVER observations are performed at several radio bands (depending on receiver availability and weather conditions) from 2.6\,GHz to 44\,GHz (11\,cm to 7\,mm wavelength). The receivers are equipped with two orthogonally polarized feeds (either circular or linear) that can deliver polarimetric observables using either native polarimeters or by connecting the SpecPol spectropolarimetric backend. Instrumental polarization is calibrated using observations of both polarized and unpolarized calibrators performed in each session and removed from the data  \cite[e.g.,][]{Krauss2003,Myserlis2018}. The polarized intensity, position angle, and polarization percentage were derived from the Stokes I, Q, and U cross-scans. For frequencies $<$10~GHz), we find a polarization degree $\Pi_R\sim5\%$ with a polarization angle $\psi_R$ between 148$^\circ$ and 188$^\circ$. The radio spectrum obtained between 2.6-43~GHz shows a smooth increase in flux towards higher frequencies, corresponding to a (hard) radio spectrum with spectral index of $a = +0.25$ (S $\propto \nu^a$). We find a continuous rotation of the polarization angle between 4.85 GHz and 10.45 GHz that suggests the presence of a Faraday screen with rotation measure of $\rm RM \sim -230 ~rad/m^2$.

The KVN observations were performed with two 21-m antennas (KVN Yonsei and Tamna) in single-dish mode. The data reduction and polarization calibration was performed following \cite{Kang2015}. We find similar results with $\Pi_R\sim6\%$ and $\psi_R$ between 16$^\circ$ and 154$^\circ$. The IRAM-30m observations were obtained and analyzed as part of the POLAMI \footnote{\url{http://polami.iaa.es/}} program \citep{agudo_polami_2018, agudo_polami_2018-1, thum_polami_2018}. The polarization degree of BL Lac was measured to be $\Pi_R = 8.08\pm1.38\%$ along position angle $\psi_R = -6.5^\circ\pm4.2^\circ$ at 1.3 mm,  and $\Pi_R = 7.27\pm0.43\%$, $\psi_R = 5.7^\circ\pm1.5^\circ$ at 3.5 mm. No circular polarization was detected ($<1.1\%$, 99\% C.I.) at 1.3 mm and $<0.6\%$ (99\% C.I.) at 3.5 mm. The SMA \citep{ho_submillimeter_2004} observation was taken within the framework of the SMA Monitoring of AGNs with Polarization (SMAPOL) program with the SMA polarimeter \citep{marrone_submillimeter_2008}. The polarized quantities are derived from the Stokes $I$, $Q$, and $U$ visibilities and calibrated with the MIR software package \footnote{\href{https://lweb.cfa.harvard.edu/~cqi/mircook.html}{https://lweb.cfa.harvard.edu/~cqi/mircook.html}}. The measurements yield $\Pi_R = 6.22\pm0.88\%$ along $\psi_R = 1.9^\circ\pm0.3^\circ$, consistent with the contemporaneous POLAMI observation.

\begin{figure}[t]
    \centering
     \includegraphics[width=0.8\textwidth]{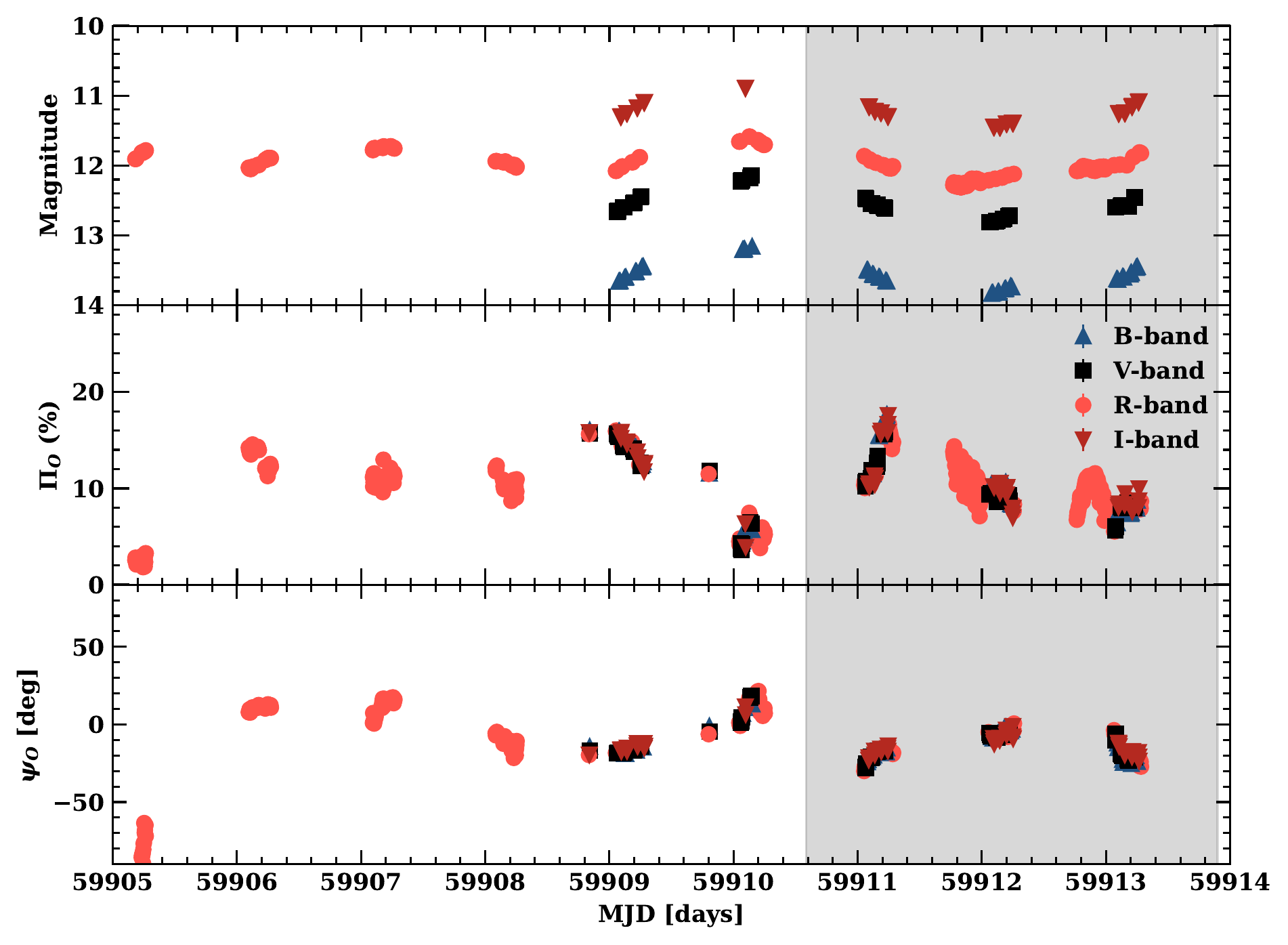}
    \caption{Optical BVRI observations of BL Lac before and during the 2022 November 27--30 {\it IXPE} pointing, showing brightness in magnitudes (top panel), polarization degree (middle panel), and polarization position angle (bottom panel). The duration of the {\it IXPE} observation is marked by the gray shaded area.}
    \label{plt:mult_obs_bllac}
\end{figure}

Optical polarization coverage was provided by the Calar Alto (Calar Alto Faint Object Spectrograph -- CAFOS, R-band), Higashi-Hiroshima Observatory (Kanata telescope) with the Hiroshima Optical and Near-InfraRed camera (HONIR -- R, J-band), the Nordic Optical Telescope (NOT) with the Alhambra Faint Object Spectrograph and Camera (ALFOSC, BVRI), the 1.8m Perkins Telescope (BVRI), and the Sierra Nevada Observatory (SNO, R-band). A detailed description of the observing strategy and data reduction of the aforementioned telescopes can be found in \cite{kawabata_new_1999,akitaya_honir_2014,hovatta_optical_2016,marscher_frequency_2021,liodakis_polarized_2022,middei_x-ray_2023}, respectively, and references therein. The observations cover the entire duration of the {\it IXPE} observation, revealing high variability of the polarization degree, from $\sim$3\% to $\sim$17\%, with EVPA fluctuating about the direction of the jet axis, $10^\circ\pm2^\circ$ \citep{weaver_kinematics_2022}. Figure \ref{plt:mult_obs_bllac} displays the optical observations, while \cref{tab:mult_obs_bllac} summarizes the radio and optical polarization results for the individual telescopes. 

\begin{figure}[t]
    \centering
     \includegraphics[width=0.7\textwidth]{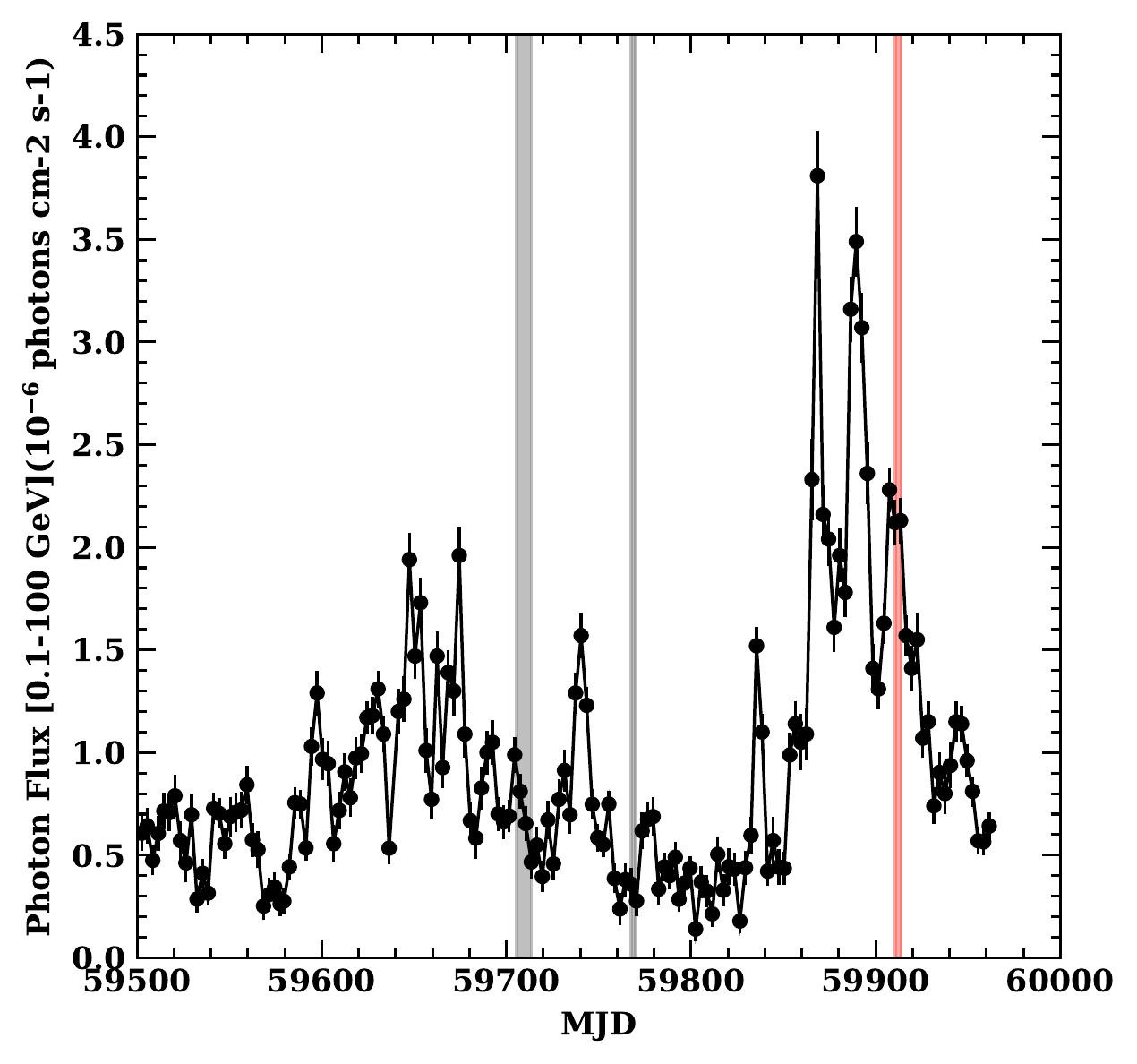}
    \caption{\textit{Fermi}-LAT 3-day cadence light curve, showing the outburst that triggered the {\it IXPE} observation. A single power-law model with a free photon index is used to determine the photon fluxes. The two gray-shaded regions demark two previous {\it IXPE} observations, while red shaded are indicates the observation discussed in this paper.}
    \label{plt:lat}
\end{figure}

The 3-day binned \textit{Fermi}-LAT $\gamma$ ray light curve of BL Lac is given in \cref{plt:lat}, extracted from the \textit{Fermi}-LAT Light Curve Repository \citep{soheila_fermi-lat_2023}, with {\it IXPE} observation times highlighted. The third {\it IXPE} observation of BL Lac, reported here, occurred during an outburst.

\begin{table}[t]
\centering
\caption{Multi-wavelength polarization observations of BL Lac during the {\rm IXPE} pointing. }
\begin{tabular}{lcccc}
\toprule
Telescope & $\rm \Pi$  (\%)& $\sigma_{\Pi}$ & $\rm \psi$ (deg.) & $\sigma_{\psi}$ \\
\midrule
KVN (22~GHz) & $5.710\pm0.066$ & -- & $16\pm5$ & --\\
KVN (43~GHz) & $6.020\pm0.138$ & -- & $6\pm1$ & --\\
KVN (86~GHz) & $4.309\pm0.621$ & -- & $172\pm7$ & --\\
KVN (129~GHz) & $6.978\pm0.493$ & -- & $154\pm4$ & --\\
POLAMI (3 mm) & $7.27\pm0.43$ & -- & $5.7\pm1.5$ & --\\
POLAMI (1.3 mm) & $8.08\pm1.38$  & -- & $-6.5\pm4.2$ & --\\
SMA (1.3 mm) & $6.22\pm0.88$  & -- & $1.9\pm0.3$ & --\\
Calar Alto \& SNO (R-band)  & $10.9\pm0.3$ &1.74 & $178\pm0.6$ &4.41 \\
Perkins (B-band) &  $10.33\pm0.22$ &3.61 & $172\pm0.6$ &11.19 \\
Perkins (V-band) &  $9.66\pm0.15$ &3.21 & $173\pm0.4$ &11.44 \\
Perkins (R-band) &  $9.28\pm0.2$ &4.10 & $175\pm0.7$ &13.51 \\
Perkins (I-band) &  $10.23\pm0.2$ &3.33 & $168\pm1$ &8.08 \\
\bottomrule
\end{tabular}
\tablecomments{The uncertainties in $\Pi$ and $\psi$ are either the  uncertainty of the measurement or, in the case of multiple measurements, the median uncertainty. $\sigma_{\Pi}$ and $\sigma_{\psi}$ show the standard deviation of the observations.}
\label{tab:mult_obs_bllac}
\end{table}

\end{document}